\documentclass{article}
\usepackage{latexsym,amssymb,amsfonts,amsmath,amsthm,graphicx}
\usepackage[utf8]{inputenc}
\usepackage{authblk}
\usepackage{setspace}
\usepackage{graphicx}
\graphicspath{ {./figures/} }
\usepackage{subcaption}
\usepackage{amsmath}
\usepackage{verbatim}
\usepackage{hyperref}
\usepackage{amsrefs}
\usepackage{float}
\usepackage{graphicx}
\usepackage{amssymb}
\usepackage{bbm}
\usepackage{bbold}
\usepackage{comment}
\usepackage{xcolor}
\usepackage{multicol}
\usepackage{neuralnetwork}
\usepackage{cancel}
\usepackage{multirow}
\usepackage{amsmath}
\usepackage{multirow}
\usepackage{amsmath}
\usepackage{amsfonts}
\usepackage{amssymb}
\usepackage{graphicx}
\usepackage{verbatim}
\usepackage{latexsym} 
\usepackage{listings}
\usepackage{tikz}
\newtheorem{theorem}{Theorem}
\newtheorem{lemma}{Lemma}
\usepackage{array}
\usepackage{makecell}
\usepackage{boldline}
\usetikzlibrary{shapes,arrows} 
\tikzstyle{decision} = [diamond, draw, fill=blue!20, 
    text width=4.5em, text badly centered, node distance=3cm, inner sep=0pt]
\tikzstyle{block} = [rectangle, draw, fill=gray!20, 
    text width=5em, text centered, rounded corners, minimum height=4em]
\tikzstyle{line} = [draw, -latex']

\usepackage{array}
\usepackage[margin=1in]{geometry}

\newcommand{\M}{\mathcal{M}}
\newcommand{\A}{\mathcal{A}}
\newcommand{\N}{\mathcal{N}}

\newcommand{\R}{\mathbb{R}}
\newcommand{\Cor}{\mathcal{C}}

\newtheorem{proposition}{Proposition}[section]
\newtheorem{corollary}{Corollary}[section]
\newtheorem{remark}{Remark}[section]

\newcommand{\s}{\vspace{2ex}}
\newcommand{\n}{\noindent}

\usepackage{amssymb}
\usepackage{graphicx}
\usepackage{mathrsfs}
\usepackage{psfrag}
\usepackage{verbatim}

\usepackage{xcolor}

\DeclareGraphicsExtensions{.eps}

\newcommand{\BE}{\mathbb{E}}

\newcommand{\T}{\mathbb{T}}
\newcommand{\deff}{\buildrel \text{def}\over =}

\begin{document}

\title{\textbf{POD-Galerkin Reduced Order Modeling of the El Ni{\~n}o-Southern Oscillation~(ENSO)}}
\author{{Yusuf Aydogdu$^1$ and 
Navaratnam Sri Namachchivaya$^1$} \vspace{0.5cm} \\
\textit{ $^1$Department of Applied Mathematics, University of Waterloo}}
\date{\today}
\maketitle

{\bf Abstract}

\vspace*{0.10cm}
Reduced order modeling~(ROM) aims to mitigate computational complexity by reducing the size of a high-dimensional state space. In this study, we demonstrate the efficiency, accuracy, and stability of proper orthogonal decomposition~(POD)-Galerkin ROM when applied to the El Ni{\~n}o Southern Oscillation model, which integrates coupled atmosphere, ocean, and sea surface temperature (SST) mechanisms in the equatorial Pacific. While POD identifies the most energetic modes of a system from simulation data, the Galerkin projection maps the governing equations onto these reduced modes to derive a simplified dynamical system. Leveraging the unique coupling properties of the model, we propose a novel approach to formulate a reduced order model derived from Galerkin projection. Our approach achieves remarkable computational efficiency, requiring only four POD modes. The results provide highly stable and accurate solutions over 95\% compared to the high-dimensional full-order model (FOM),  highlighting the potential of POD-Galerkin reduction for efficient and accurate climate simulations.

\section{Introduction}
Simulations of complex fluid dynamics problems or climate models take weeks to complete even when run parallel in state-of-the-art supercomputers. Given limitations in computational resources or simulation setting requirements, there is a need for cost-efficient and accurate algorithms. To overcome these difficulties, reduced order models aim to reduce the computational complexity and present high-dimensional problems~(usually PDEs) with reduced order equations (ODEs). Although PDEs and their time series solutions are infinite-dimensional, functional data sets can often be effectively represented using a finite ($n \ll \infty$) number of basis functions. Proper orthogonal decomposition (POD) provides an optimal set of basis functions, enabling the construction of a low-dimensional subspace. Projecting the governing PDEs onto this subspace yields a set of reduced-dimensional equations.\newline

In the analysis of functional data of random processes, each observation is a curve on a  functional space assumed to be sampled from a stochastic process $X\in L^{2}[0,T]$ - the Hilbert space of square integrable functions on the $[0,T]$ with the usual inner product $\big\langle \varphi,\psi  \big\rangle\deff \int_{0}^{T} \varphi(s) \psi(s) ds $ for any two functions $\varphi,\psi \in L^{2}[0,T]$. As the primary theoretical foundation of POD, 
The Karhunen-Lo\'eve expansion  of $X$, under certain technical conditions, is assumed to exist such that 
$$
X(t)=\mu_{_{X}}(t)+\sum_{j=1}^{\infty}\xi_{j}\varphi_{j}(t),
$$
where the mean function $\mu_{_{X}}(t)\deff\BE\big[X(t)\big]$ and the basis functions $\varphi_{j}(t)$ are the orthonormal eigenfunctions of the covariance {\em kernel} $\Gamma_{_{X}}(s,t)=\text{Cov}\big[X(s),X(t)\big]$ \cites{Karhunen1947, Loève1978}.
The eigenvalues corresponding to $\varphi_{j}(t)\in L^{2}[0,T]$ satisfy
\begin{equation*}
\begin{aligned}
&\int_{0}^{T}\Gamma_{_{X}}(s,t) \varphi_{j}(s)ds =\lambda_{j} \varphi_{j}(t) 
\quad \text{and}\quad 
&\Gamma_{_{X}}(s,t) =\sum_{j=1}^{\infty} \lambda_{j} \varphi_{j}(t)  \varphi_{j}(s).
\end{aligned}
\end{equation*}
The random coefficients $\xi_{j}$ are given by the projection of $X(t)-\mu_{_{X}}(t)$ in the direction of the $j^{\text{th}}$ eigenfunction $\varphi_{j}$, that is,
$$\xi_{j}\deff \big\langle X(t)-\mu_{_{X}}(t),  \varphi_{j} \big\rangle.$$ Furthermore, $\big\{\xi_{j}\big\}$ are uncorrelated sequences of random variables with zero mean and variance 
$\lambda_{j}$. Since the process $X\in L^{2}[0,T]$, we have $\sum_{j=1}^{\infty} \lambda_{j}<\infty$. Extending this method by considering {\em spatial velocity correlations}  was an essential ingredient first introduced to the fluids community by Lumley~\cite{Lumley1970}.
The fluid flows are assumed to be turbulent and time stationary. Under these conditions, ensemble averages were taken to form the spatial velocity correlations.
The orthonormal eigenfunctions of the velocity correlations were then used as a rational and quantitative basis of identifying coherent structures.\newline

Although solutions are defined on the interval $[0,T]$ they are seldom observed there; instead they are observed on a discrete subset of points. 
The {POD} method of snapshots, where the data are taken at different times as a way to efficiently determine the $n<<\infty$ basis functions, was first suggested by Sirovich~\cite{Sirovich1987} to study fluid dynamic applications for extracting coherent or dominant structures.
The only requirement is that the snapshots are linearly independent~(or data generated by ergodic dynamical systems).  Given a function $w(t,x)$,  the POD seeks a decomposition of the form
\begin{equation*}
   w(t,x) \approx  \sum_{\alpha = 1}^{n} a_{\alpha} (t) \psi_{\alpha} (x) 
\end{equation*}
with  spatial basis functions $\psi_{\alpha}(x)$ (also called modes), orthogonal in the spatial domain, and temporal coefficients $a_{\alpha}(t)$, capturing the time evolution of the function. $n$ is the number of retained modes, chosen based on the desired level of approximation.
 \\

An open question is how a given functional data set can effectively be spanned by finite basis functions \textit{without the assumption of ergodicity.} 
There is little in the current literature that provide sufficient answer to this question with a notable exception of Djouadi and Sahyoun~\cite{Djouadi2012}.
Their work explicitly investigates the theoretical properties of dimension reduction techniques within a functional analytic framework. That conceptual framework is extended in this paper. The techniques presented  in ~\cite{Djouadi2012} circumvent one of the confounding issues that the snapshots are linearly independent. 
In addition, we develop the theoretical properties of the techniques within a functional analytic framework to identify an optimal set of basis functions for vector valued coupled nonlinear PDEs in one spacial dimension. These results give an efficient way to construct the basis functions, which will be used for model reduction.\newline

Along with POD, Galerkin approximations are used to build weak solutions of the PDEs by constructing solutions of certain finite-dimensional~($n$) approximations to the PDEs.
Many ROM techniques in fluid mechanics arise from the POD-Galerkin projection approach \cites{Aubry1988,Sirovich1987}.
 Suppose we are given the PDE in the following form \begin{equation}
    \frac{\partial w}{\partial t} (t,x) =  \A  w(t,x) + \N(w,w) (t,x) 
\end{equation}
The Galerkin projection reduces the PDE model to a set of reduced-order equations (ODEs) by projecting the governing equations onto a subspace spanned by the reduced basis functions, resulting in the following form:
\begin{equation*}
    \dot{a}_{\gamma}(t) = \sum_{\alpha=1}^{n} a_{\alpha}(t) \langle \A \psi_{\alpha} , \psi_{\gamma} \rangle  + \sum_{\alpha, \beta =1}^{n} a_{\alpha} (t) a_{\beta} (t) \langle \N (\psi_{\alpha}, \psi_{\beta}) , \psi_{\gamma} \rangle \hspace{1cm} \gamma = 1, ... , n
\end{equation*}
The main application in this paper is a minimal ocean and atmosphere circulation model -- linked by exchanges at the air-sea interface - that captures the evolution of  El Ni{\~n}o Southern Oscillation (ENSO) patterns. This modeling framework was originally developed by Majda and co-workers~\cites{Chen2018, Thual2016} and is physically consistent and amenable to detailed analysis.
Due to the special coupling and boundary properties of the model, we adapt the POD-Galerkin method and introduce a novel approach to obtain stable and accurate results. We show that a full order coupled ENSO model can be approximated using a few POD modes and reduced order equations.  Our approach provides efficient, stable, and accurate results, showcasing its potential for advancing reduced-order modeling in this domain.\newline

This paper is organized as follows: we introduce the coupled ENSO mechanism formed by a 5-dimensional PDE model and discuss some of its properties in the next section. We give an overview of POD technique in Section \eqref{sc:pod} and along with the main theorem we also show how to construct POD bases using snapshots and correlation matrices. We provide POD modes and time coefficients of the nonlinear ENSO model in this section. In Section \eqref{sc:pod-enso}, we introduce POD method combined with Galerkin projection to obtain reduced order equations associated with the ENSO model. Finally, we show our results in Section \eqref{sc:results} and discuss the advantages of our proposed method and future directions in Section \eqref{sc:conc}.

\section{Model Description : El Ni{\~n}o Southern Oscillations}\label{sec:Model}

The El Niño Southern Oscillation (ENSO) is one of the most influential climate phenomena, affecting weather patterns across the globe. Originating in the tropical Pacific, this system involves complex ocean-atmosphere interactions. This phenomenon entails the interplay between the ocean and atmosphere and significantly affects global weather patterns. Initial observations of low-latitude climatic anomalies by Bjerknes \cite{Bjerknes1969} in the 1960s established the link between sea surface temperature (SST) and equatorial atmospheric circulations, connecting El Ni{\~n}o to the Southern Oscillation. Later, Matsuno and Gill \cite{Gill1980} used simplified one-layer shallow-water models with damping to analyze atmospheric responses via equatorial Kelvin and Rossby waves. Since the 1970s, experimental forecasts of ENSO have been conducted. Among the intermediate atmosphere-ocean coupled models, the Cane-Zebiak model \cite{CZ1987} has been particularly influential, accurately capturing the key dynamics of the coupled system. Additionally, Neelin, Dijkstra, and their colleagues proposed a comprehensive deterministic ENSO theory \cites{Dijkstra2006, Neelin1998}, thoroughly analyzing the system components. \\

Coupled ENSO models primarily focus on the upper layer of the tropical Pacific Ocean. This layer drives energy and momentum exchange with the tropopause. The deep ocean, in contrast, is considered dynamically inactive. The upper ocean of interest is a very long strip of water extending from 124E to 80W and from 29S to 29N with: i) a meridional (north-south) width of a few hundred kilometers, ii) a zonal (east-west) extension of about 10,000 kilometers and iii) an average thermocline depth (downwards from the sea surface) of about 120 meters, which is a relatively thin ocean layer above the deep cold waters.\\

In this paper,  we present the nonlinear ENSO model developed by Majda and colleagues \cites{Chen2018, Thual2016}~(Majda's model), where the ocean and atmosphere are both modeled as non-dissipative linear shallow water equations, with a dissipative nonlinear thermodynamics budget. Similar to the Zebiak-Cane model, it is an anomaly model, where the anomalies are calculated relative to a mean annual cycle specified from observations, and thus all state variables in subsequent sections are anomalies from an equilibrium state and are nondimensional. Building on the foundation of the Cane-Zebiak model, Majda's model addresses the limitations of linearity and determinism, making it a more versatile tool for studying the complexities of ENSO phenomena. This model provides a more realistic framework for studying ENSO variability, extremes, and predictability in the presence of noise, although we use deterministic version in this paper. Such models can then be further simplified by meridional truncation that reduces the dimensionality of the problem by approximating the solution in the meridional (north-south) direction using a predefined set of modes.

We utilize this nonlinear ENSO model  to implement the proposed POD-Galerkin method for high-dimensional coupled problems. The dissipative steady atmosphere model is driven by atmospheric Kelvin $K^{A}(t,x)$ and Rossby $R^{A}(t,x)$ waves, respectively. Moreover, a dissipative shallow water ocean model forced by atmospheric waves is composed of oceanic Kelvin $K^{O}(t,x)$ and Rossby $R^{O}(t,x)$ waves. The final component of the coupled model is the sea surface temperature (SST) model $T(t,x)$ that characterizes the temperature anomalies.\\

Let \(\Omega:=\{x \in [0,L] \}\) denote the extent of the  equatorial Pacific Ocean~(the physical space between Australia and South America) 
and the time domain be defined with with a terminal time $T$ as \( \T:=\{t \in [0,T] \}\). Then the nonlinear coupled PDEs in five  variables which model the interaction of Kelvin and Rossby waves with SST anomalies, defined as
$$u(t,x)\deff[K^{A}(t,x), R^{A}(t,x) , K^{O}(t,x) , R^{O}(t,x) , T(t,x)]^{T}$$ and can be written in differential equation form as 
\begin{equation}
    \M \frac{\partial u}{\partial t} (t,x) =  \A (\kappa) u(t,x) +  
    \N(u,u,\mu) (t,x),
     \quad \text{in} \;\;  \T\times \Omega \label{eq:Coupled_Model} 
\end{equation}
where $\M$ is a $5 \times 5$ constant matrix with diagonals $[0,0,1,1,1]$, the linear operator $A(\kappa)$ defined as
\begin{equation}
 \label{eq:A_opertor} 
\begin{aligned}
    \A (\kappa) &= \begin{bmatrix}
       - \frac{\partial }{\partial x} - \gamma & 0 & 0 & 0 & \chi_{A}  \alpha_{q}  (2 -  2\overline{Q})^{-1}  \\
        0 & \frac{1}{3} \frac{\partial }{\partial x} - \gamma & 0 & 0 & \chi_{A} \alpha_{q}  (3 - 3 \overline{Q})^{-1}   \\
         \chi_{O} \frac{c \kappa}{2} &  -\chi_{O} \frac{c \kappa}{2} & -c \frac{\partial}{\partial x} - \delta & 0 & 0  \\
        -\chi_{O} \frac{c \kappa}{3} &  \chi_{O} \frac{c \kappa}{3} & 0 & \frac{c}{3} \frac{\partial}{\partial x} - \delta  & 0   \\
        0 & 0 & c \eta (x) & c \eta (x) & -c \xi \alpha_q    
    \end{bmatrix}
    \end{aligned}
\end{equation}
depends on a parameter $\kappa$, the wind stress coefficient, and nonlinearity $\N(u,u, \mu)$ with advection coefficient $\mu$, is defined by
$$
\N(u,u, \mu) \deff  \big[ 0, 0, 0 , 0 , -\mu \frac{\partial }{\partial x}[(K^{O}(t,x) - R^{O}(t,x))T(t,x)\big]^{T}.
$$

In equation~\eqref{eq:Coupled_Model},
 small damping terms \( \gamma,   \delta\) are introduced to guarantee a unique solution, \(E_q\) is latent heat,  \(\xi\) is the latent heating exchange coefficient,
\(\Bar{Q}\) is a constant representing mean vertical moisture gradient, \(\eta(x)\) is the profile of thermocline feedback and  the projection coefficients are defined as 
$$\chi_{A}=\int_{-\infty}^{\infty} \phi_{0}(y)\psi_{0}(Y) \, dy \;\;\text{and} \;\;\chi_{O}=\int_{-\infty}^{\infty} \psi_{0}(Y)\phi_{0}(y) \, dY,$$
where \(\phi_{0}\)  and  \(\psi_{0}\) denote the atmospheric and oceanic parabolic cylinder functions as a function in meridional position $y$ and $Y = y/ \sqrt{c}$.

The initial conditions for the oceanic variables of~\eqref{eq:Coupled_Model} are given by
\begin{equation}\label{5D_model_IC}
    u_3(x, 0)=f_3(x), \;\;u_4(x, 0)=f_4(x)\;\;\text{and}\;\; u_5(x, 0)=f_5(x).
\end{equation}
Periodic boundary conditions that were adopted for the atmosphere model in~\cite{Chen2018} are modified to reflect the inclusion of the small damping term $\gamma$.
Boundary conditions for the atmosphere variables that are confined in the ocean domain: 
\begin{equation}
K^{A} (t,0) = e^{-\gamma (L_{A} - L_{O})} K^{A}(t,L_{O}) \hspace{0.35cm} \text{and} \hspace{0.235cm}  R^{A} (t,0) = e^{3 \gamma (L_{A} - L_{O})} R^{A}(t,L_{O})  \label{eq:BC1}
\end{equation} 
where $L_A$ and $L_O$ represent the length of the equatorial belt and Pacific ocean, respectively. Having defined the ocean domain, the next step is to specify the boundary conditions that govern wave interactions and energy exchange as the following:
\begin{equation}
    K^{O} (t,0) = r_{W}R^{O} (t,0) \hspace{0.35cm} \text{and} \hspace{0.35cm} R^{O} (t,L_{O}) = r_{E} K^{O} (t,L_{O}) \label{eq:BC2}
\end{equation}
where $x=0$ and $x=L_O$ represent the western and eastern parts of the ocean, respectively. The unidirectional Kelvin and Rossby oceanic equatorial waves
are coupled through the boundary condition~\eqref{eq:BC2}, where  the oceanic waves are reflected with coupling coefficients $r_{W}$ that represents partial loss of energy in the western Pacific boundary and $r_{E}$ that represents partial loss of energy due to the north-south propagation of the coastal Kelvin
waves along the eastern Pacific boundary.

Equation~\eqref{eq:Coupled_Model} consists of a simple nonlinear advection $-\mu \frac{\partial }{\partial x}[(K^{O}(t,x) - R^{O}(t,x))T(t,x)]$ in the SST equation.
We assume no heat flux at the eastern boundary, as specified by the following boundary conditions for $T(t,x)$:
\begin{equation*}
     \frac{\partial T}{\partial x} |_{x=L_O} = 0
\end{equation*}

Suppose $\{u(t,x): 0 \le t \le T, x \in \Omega\}$ is the solution of the coupled PDE~\eqref{eq:Coupled_Model}, defined on some domain $\Omega$ and taking values in some Euclidean space $\R^5$.  Assume the solution $u(t,.) \in L^2([0,T], H)$ where $H$ is some Hilbert space of functions $\Omega \to \R^5$ such as an $L^2$~(square-integrable vector functions \(u\) on \(\Omega\)) or \(H^1(D) \subset L^2(D)\) which denotes the Sobolev space of \(u\) of order one that satisfies the boundary conditions \eqref{eq:BC1} and \eqref{eq:BC2}.

\section{Proper Orthogonal Decomposition}
\label{sc:pod}
\noindent
In this section, we develop the theoretical properties of the  POD techniques within a functional analytic framework~(originally introduced by Djouadi and Sahyoun~\cite{Djouadi2012}) to identify theoptimal set of basis functions for vector valued coupled nonlinear PDEs. These results give an efficient way to construct the basis functions, which will be used for model reduction. We thank Peter Baxendale for permission to copy the material in \cite{Baxendale2024}. \\

Suppose $\{w(t,x): 0 \le t \le T, x \in \Omega\}$ is the solution of some PDE, defined on some domain $\Omega$ and taking values in some Euclidean space $\R^d$.  Assume the solution $w(t,.) \in L^2([0,T], H)$, where $H$ is some Hilbert space of functions $\Omega \to \R^d$ such as an $L^2$ or Sobolev space.  In order to construct a reduced order model (ROM) we first look for a finite-dimensional subspace $H_0$ of $H$ and a function $\widetilde{w}: [0,T] \to H_0$ taking values in $H_0$ so that the $L^2$ difference 
        $$
        \int_0^T \|w(t,\cdot) - \widetilde{w}(t)\|_H^2dt
        $$
is small.  The choice of $H_0$, equivalently the choice of a set of ``POD modes'' to form a basis for $H_0$, is made using only observations of the solution $w(t,x)$.  Information about the PDE will be used later to find an ODE satisfied approximately by the finite-dimensional process $\{\widetilde{w}(t): 0 \le t \le T\}$.

\s

Instead of continuous time observations, suppose we have snapshots $\{w(x,t_i); 1 \le i \le M \}$ of the function $w(t,x)$ taken at times $t_i$ uniformly spread through the interval $[0,T]$.   
  That is, we have elements $w(t_i) = w(t_i,\cdot) \in H$ for $1 \le i \le M$, and we look for a finite-dimensional subspace $H_0$ of $H$ and a sequence $\widetilde{w}_1, \ldots, \widetilde{w}_M$ taking values in $H_0$ so that the $L^2$ difference 
        $$
        \sum_{i=1}^M \|w(t_i)-\widetilde{w}_i\|_H^2
        $$
is small.  More precisely, given $n \ge 1$ let  ${\cal S}_n$ be the set of all functions $F: \{1,2,\ldots,M\} \to H$ of the form 
   $$
   F_i = \sum_{\alpha= 1}^n a^\alpha_i\psi_\alpha
   $$
where all the vectors $a^\alpha$ are in $\R^M$ and all the $\psi_\alpha$ are in $H$.  Thus $F_i$ takes values in the linear span $H_0 = \mbox{span}\{\psi_1,\ldots,\psi_n\}$ which has dimension at most $n$.  Define
   \begin{equation} \label{mu}
    \mu_n = \inf_{F \in {\cal S}_n} \left( \sum_{i=1}^M \|w(t_i)- F_i\|_H^2\right).
   \end{equation}

\s

\begin{theorem} \label{thm opt}   Given the snapshots $w(t_1),\ldots,w(t_M)$ define the $M \times M$ matrix $C$ by
     \begin{equation} \label{C}
         C_{ij} = \langle w(t_i),w(t_j) \rangle_H, \qquad 1 \le i,j \le M.
      \end{equation}
The matrix $C$ is symmetric and non-negative, with eigenvalues $\lambda_1\ge \lambda_2 \ge \cdots \ge 0$ (counted with multiplicities) and corresponding orthonormal eigenvectors $\gamma^1, \gamma^2, \ldots$ in $\R^M$.   Then the minimization problem \eqref{mu} has solution $\mu_n = \sum_{\alpha > n} \lambda_\alpha$ and the optimal $F$ is given by $F_i = \sum_{\alpha= 1}^n a^\alpha_i\psi_\alpha$ where $\{\psi_1, \ldots, \psi_n\}$ is orthonormal and is given by  
     $$
     \psi_\alpha = \frac{1}{\sqrt{\lambda_\alpha}}\sum_{i=1}^M \gamma^\alpha_i w(t_i), \qquad 1 \le \alpha \le n,
     $$
and 
    $$
     a^\alpha_i = \sqrt{\lambda_\alpha} \gamma^\alpha_i, \qquad 1 \le i\le M, \, 1 \le \alpha \le n.
     $$
\end{theorem}              

The proof of Theorem \ref{thm opt} is given in the Appendix A.

   \section{POD-Galerkin Method Applied to the ENSO Model}
   \label{sc:pod-enso}
   
 In this section, we apply POD techniques presented in Section~\ref{sc:pod} to the ENSO model described in Section~\ref{sec:Model}.
 Suppose $\{u(t,x): 0 \le t \le T, x \in \Omega\}$ is the solution of the coupled PDE~\eqref{eq:Coupled_Model}, defined on some domain $\Omega$ and taking values in some Euclidean space $\R^5$. 
In~\eqref{eq:Coupled_Model}, the equatorial atmosphere is modeled by a steady-state linear equation forced by heat from the equatorial Pacific Ocean surface, which is proportional to the SST $T(t,x)$. 
 Hence, the first two components $K^A(t,x)$ and $R^{A}(t,x)$ of $u(t,x)$ are driven by $T(t,x)$. 
For each fixed time $t$ the first two equations along with the boundary conditions on $K^A(t,x)$ and $R^{A}(t,x)$ the functions $x\mapsto K^A(t,x)$ and $x\mapsto R^A(t,x)$ 
 are determined by the function $x\mapsto T(t,x)$~(See equations~\eqref{eq:atmosM-truncate-Ksol} and~\eqref{eq:atmosM-truncate-Rsol} in Appendix B for explicit representation of $K^A(t,x)$ and $R^{A}(t,x)$ in terms of SST $T(t,x)$). Thus there are linear
transformations $\Phi_1$ and $\Phi_2$ sending functions of $x$ to functions of $x$ such that for each $t$ we
have 
 $$
 K^A(t,\cdot)= \Phi_1(T(t,\cdot)),\quad\text{and}\quad  K^A(t,\cdot)=\Phi_2(T(t,\cdot)).
 $$
 Therefore, every solution $u(t,x)$ of  \eqref{eq:Coupled_Model} satisfying the 
 reflection boundary conditions that couples $K^{O}(t,x)$ and $R^{O}(t,x)$  is of the form 
\begin{equation}
    u(t,x) = [\Phi_1(T(.,t))(x), \Phi_2(T(.,t))(x) , K^{O}(t,x) , R^{O}(t,x) , T(t,x)] \label{eq:soln}
\end{equation}

\begin{figure}[H]
     \centering
     \begin{subfigure}[b]{0.45\textwidth}
         \centering
         \includegraphics[scale=0.225]{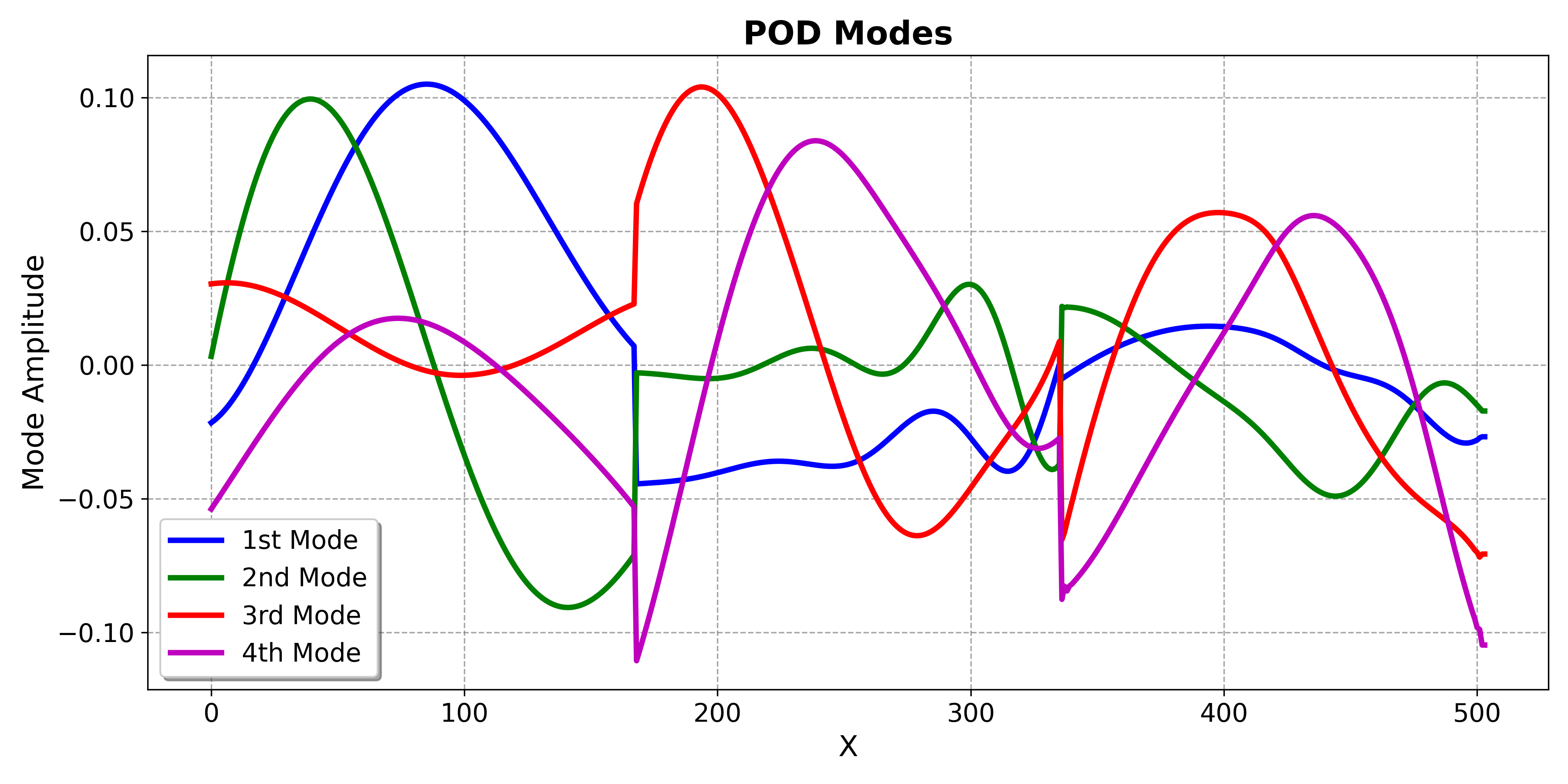}
         \caption{}
     \end{subfigure}
     \hfill
     \begin{subfigure}[b]{0.45\textwidth}
         \centering
         \includegraphics[scale=0.225]{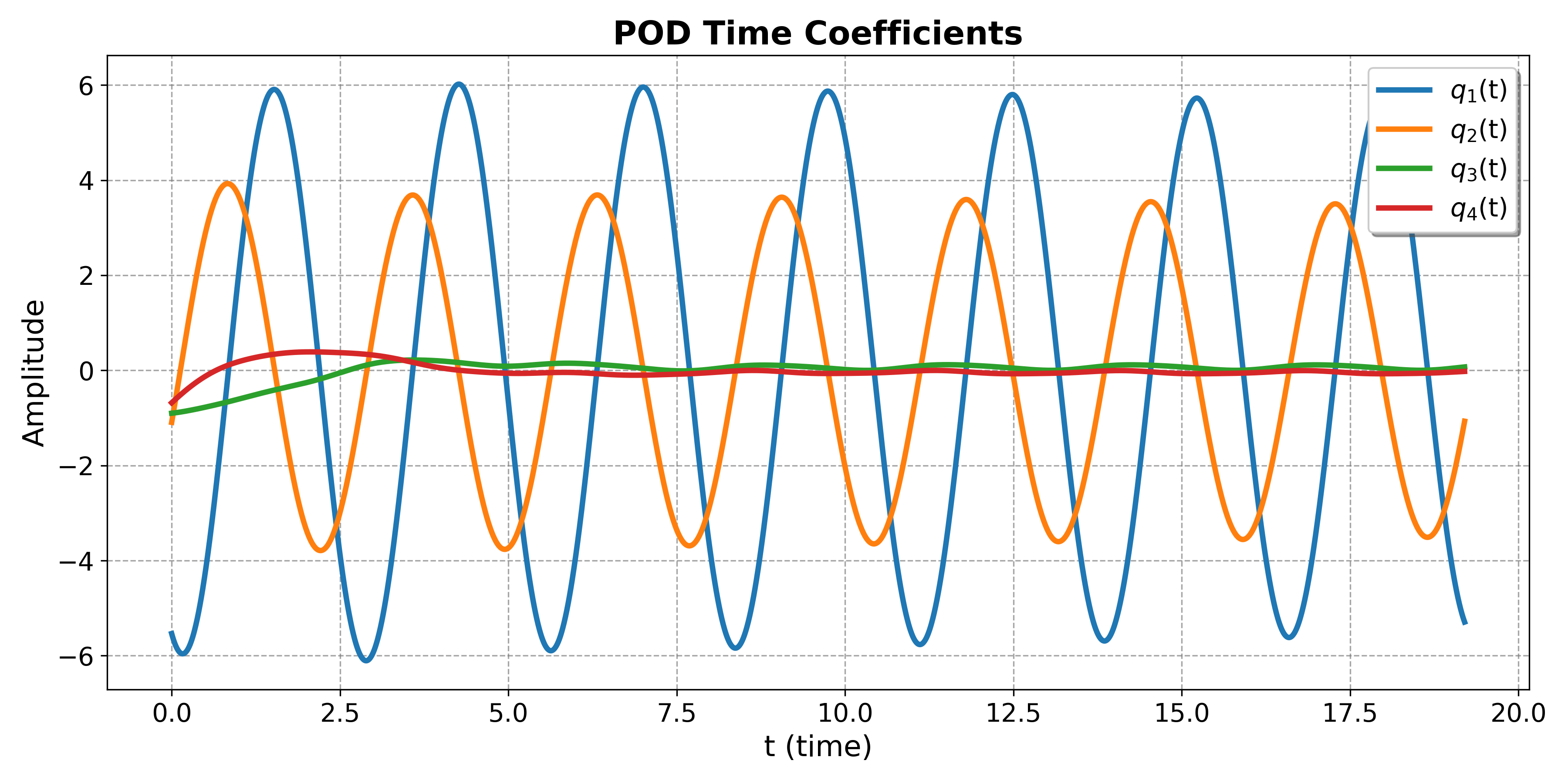}
         \caption{}
     \end{subfigure}
     \caption{ (a) POD Modes $\tilde{\Psi}_{\alpha}(x)$ and (b) POD time coefficients $q_{\alpha}(t)$ }
     \label{fig:POD1}
\end{figure}

Assume the solution $u \in L^2([0,T], H)$ where $H$ is some Hilbert space of functions $\Omega \to \R^5$  such as an $L^2$ or Sobolev space.
 Due to the functional dependance of $K^A(t,x)$ and $R^{A}(t,x)$ on  $T(t,x)$ as shown in~\eqref{eq:soln}, for the ENSO system, we concentrate on the last three components of $u(t,x)$, that is, functions
\begin{equation}
   \overline{u}(t,x) = \big( K^{O}(t,x) , R^{O}(t,x) , T(t,x)\big) .\label{eq:solnbar}
\end{equation}
 Recall that the solution \( u(t,x) \) satisfies \( u(t,\cdot) \in H \) for all \( t \), where \( H \) is a Hilbert space consisting of functions \([0,L] \to \mathbb{R}^5\). Define the new Hilbert space
\[
\hat{H} = \{ v : [0,L] \to \mathbb{R}^3 : (0,0,v) \in H \}
\]
with the inner product
\[
\langle v, w \rangle_{\hat{H}} = \langle (0,0,v), (0,0,w) \rangle_{H}, \quad v, w \in \hat{H}.
\]
Here if \( v(x) = (K^O(t,\cdot), R^O(t,\cdot), T(t,\cdot)) \) then \( (0,0,v) : [0,L] \to \mathbb{R}^5 \) is the function
\[
x \mapsto (0,0,K^O(t,\cdot), R^O(t,\cdot), T(t,\cdot)).
\]
\textbf{Remark 4.1. :} \textit{If the Hilbert space \( H \) is the product of Hilbert spaces \( H_1 \times H_2 \times H_3 \times H_4 \times H_5 \), so that \( u \in H \) can be written}
\[
u(x) = (u_1(x), u_2(x), u_3(x), u_4(x), u_5(x))
\]
\textit{with \( u_i \in H_i \) for \( i = 1, \dots, 5 \) and}
\[
\langle u, v \rangle_H = \sum_{i=1}^{5} \langle u_i, v_i \rangle_{H_i}, \quad u, v \in H
\]
\textit{then \( \hat{H} = H_3 \times H_4 \times H_5 \) .} Therefore, for each fixed time $t$ the mapping $x\mapsto  \big( K^{O}(t,x) , R^{O}(t,x) , T(t,x)\big)$ is in $\hat{H}$. It should be noted that we define $H = H_{KA} \times H_{RA} \times H_{KO} \times  H_{RO} \times H_{T}$ and $\hat{H} =  H_{KO} \times  H_{RO} \times H_{T} $ in the PDE model \eqref{eq:Coupled_Model}, respectively. Considering this, we can define full snapshots 
 
\begin{equation*}
    u_i (x) = (\Phi_1(T(t_i,.))(x), \Phi_2(T(t_i,.))(x) ,K^{O}(t_i,x), R^{O}(t_i,x),T(t_i,x)) \hspace{1cm} 1 \leq i \leq M
\end{equation*}
and reduced snapshots as
\begin{equation}
    \overline{u}_i (x) = (K^{O}(t_i,x), R^{O}(t_i,x),T(t_i,x)) \hspace{1cm} 1 \leq i \leq M \label{eq:reduced_snapshots}
\end{equation}
where $\overline{u}_i(x) \in \hat{H} $. 
The new Hilbert space $\hat{H}$ is a product of Hilbert spaces $H_{KO}\times H_{RO}\times H_{T}$.
We can form an $M \times M$ matrix using the reduced snapshots such as
\begin{equation*}
\begin{aligned} 
    \Cor_{ij} &= \frac{1}{M} \langle \overline{u}_i , \overline{u}_j \rangle_{_{\hat{H}}} \\
    &=  \frac{1}{M} \Big ( \langle K^O_i(x) , K^O_j(x) \rangle_{_{H_{KO}}} + \langle R^O_i(x) , R^O_j(x) \rangle_{_{H_{RO}}} + \langle T_i(x) , T_j(x) \rangle_{_{H_{T}}}  \Big ) \hspace{1cm} 1 \leq i , j \leq M  
    \end{aligned}
\end{equation*} 
Using  the reduced snapshots given in \eqref{eq:reduced_snapshots} and the eigenvectors $\gamma^{\alpha}$ and eigenvalues $\lambda_{\alpha}$ of matrix $\Cor$, we can define the orthonormal modes  by (see Theorem \eqref{thm opt})
\begin{equation}
    \tilde{\Psi}_{\alpha}(x) = \frac{1}{\sqrt{ \lambda_{\alpha}}} \sum_{i=1}^{M} \gamma^{\alpha}_{i} \overline{u}_{i}(x)  \hspace{1cm} \text{and} \hspace{1cm} \hat{\Psi}_{\alpha}(x) = \frac{1}{\sqrt{ \lambda_{\alpha}}} \sum_{i=1}^{M} \gamma^{\alpha}_{i} u_{i}(x)
\end{equation}
where  $\tilde{\Psi}(x)$  and $\hat{\Psi}(x)$ are constructed using reduced and full snapshots, respectively. Figure \eqref{fig:POD1} illustrates the POD modes $\tilde{\Psi}_{\alpha}(x)$ and POD time coefficients $q_{\alpha}(t)$. It is worth pointing out that POD modes $\tilde{\Psi}(x)$  and $\hat{\Psi}(x)$ remain invariant regardless of initial conditions for the PDE model \eqref{eq:Coupled_Model}. As the full order PDE model given in Equation \eqref{eq:Coupled_Model} is discretized using 168 spatial nodes, each POD mode consists of 504 elements, combining the corresponding modes for $K^O(t,x)$, $R^O (t,x)$ and $T(t,x)$, respectively. We observe that four modes are highly stable and sufficient to capture the main features of the full-order model $u(t,x)$ while the first two modes are the most dominant, effectively capturing the primary dynamics of the full-order model. The reconstruction results are presented in the next section to show the accuracy of the method.
After finding optimal POD bases $\{ \tilde{\Psi}_1, ... ,\tilde{\Psi}_n \}$ and $\{ \hat{\Psi}_1, ... ,\hat{\Psi}_n \}$, we will solve an approximate problem on a finite dimensional subspace of functions, and then pass to a limit. Hence one can project PDE \eqref{eq:Coupled_Model}  onto these POD modes using Galerkin projection to find set of ODEs for the $\R^n$ valued process $t \rightarrow (a_1(t), ... , a_n(t) )$ so that the solution given by $u_n (t,x)$ stays close to the solution $u(t,x)$.  The steady state properties of first two components of model \eqref{eq:Coupled_Model} suggest that the time evolution of the system is controlled by the evolution of the last three components $K^{O}(t,x)$, $R^{O}(t,x)$ and $T(t,x)$ satisfying 
\begin{align}
   \frac{\partial}{\partial t} \overline{u}(t,x) &=  \begin{bmatrix}
        \frac{\partial K^{O}}{\partial t} (t,x) &
         \frac{\partial R^{O}}{\partial t} (t,x) &
          \frac{\partial T}{\partial t} (t,x) 
    \end{bmatrix}^T \nonumber \\  
    &= \A_0 u (t,x) + \N_0 (u,u) (t,x) \label{eq:_ReducedPDE}
\end{align}
where $u(t,x)$ is given by \eqref{eq:soln} and $\A_0$ and $\N_0 (u,u)$ denote the {\em last three rows} of $\A$ and $\N(u,u)$, respectively. Replacing $u(t,x)$ and $\overline{u}(t,x)$ with 
\begin{equation}
\label{appximation}
u_n (t,x)=\sum_{\alpha=1}^{n}a_{\alpha}(t)\hat{\Psi}_{\alpha}(x)\quad\text{and}\quad\overline{u}_n(t,x)=\sum_{\alpha=1}^{n}a_{\alpha}(t)\tilde{\Psi}_{\alpha}(x)
\end{equation} 
in \eqref{eq:_ReducedPDE} results in 
\begin{equation}
    \sum_{\alpha=1}^{n} \dot{a}_{\alpha}(t) \tilde{\Psi}_{\alpha} = \sum_{\alpha=1}^{n}  a_{\alpha}(t) \A_0 \hat{\Psi}_{\alpha} + \sum_{\alpha, \beta =1}^n a_{\alpha}(t) a_{\beta}(t) \N_0 (\hat{\Psi}_{\alpha}, \hat{\Psi}_{\beta})  
\end{equation}
The PDEs for $K^{O}$ and $R^{O}$ are coupled through the boundary conditions~\eqref{eq:BC2}, that is,
$K^{O} (t,0) = r_{W}R^{O} (t,0)$ and $R^{O} (t,L_{O}) = r_{E} K^{O} (t,L_{O})$.
Hence,  due to this coupling, the amplitudes $\big({a}_{1}(t), {a}_{2}(t),\cdots {a}_{n}(t)\big)$ in~\eqref{appximation} are common for all the POD modes, as described in Remark 4.2 below. Taking the inner product of each side with the function $\tilde{\Psi}_{\gamma}$ such that 
\begin{equation}
    \sum_{\alpha=1}^{n} \dot{a}_{\alpha}(t) \langle \tilde{\Psi}_{\alpha},\tilde{\Psi}_{\gamma} \rangle_{\hat{H}}  = \sum_{\alpha=1}^{n}  a_{\alpha}(t) \langle \A_0 \hat{\Psi}_{\alpha}, \tilde{\Psi}_{\gamma} \rangle_{\hat{H}} + \sum_{\alpha, \beta =1}^n a_{\alpha}(t) a_{\beta}(t) \langle \N_0 (\hat{\Psi}_{\alpha}, \hat{\Psi}_{\beta}), \tilde{\Psi}_{\gamma}  \rangle_{\hat{H}} 
\end{equation}
These assumptions suggest the following reduced order equations: 
\begin{equation}
    \dot{a}_{\gamma}(t) = \sum_{\alpha=1}^{n} a_{\alpha}(t) \langle \A_0 \hat{\Psi}_{\alpha} , \tilde{\Psi}_{\gamma} \rangle_{\hat{H}}  + \sum_{\alpha, \beta =1}^{n} a_{\alpha} (t) a_{\beta} (t) \langle \N_0 (\hat{\Psi}_{\alpha}, \hat{\Psi}_{\beta}) , \tilde{\Psi}_{\gamma} \rangle_{\hat{H}} \hspace{1cm} \gamma = 1, ... , n \label{eq:Reduced_ODEs}
\end{equation}
Since
\begin{equation}
\langle \tilde{\Psi}_{\alpha},\tilde{\Psi}_{\gamma} \rangle_{\hat{H}} =  \begin{cases}
        1 & \alpha = \gamma \\ 
        0 & \alpha \neq \gamma  
    \end{cases}
\end{equation}
Equation \eqref{eq:Reduced_ODEs} represents the reduced order model (ODEs) of the full order PDE model \eqref{eq:Coupled_Model}. Similar to POD analysis, we observe that four modes ($n=4$) is sufficient to capture the main properties using the POD-Galerkin method. To simulate the ODEs, the initial conditions of PDE are projected to determine the corresponding initial conditions given  in \eqref{5D_model_IC} for the reduced-order model using  $\langle (u_3(0,x), u_4 (0,x), u_5(0,x)), \tilde{\Psi}_{\alpha} \rangle $ . \\
\begin{figure}[H]
     \centering
     \begin{subfigure}[b]{0.45\textwidth}
         \centering
         \includegraphics[scale=0.225]{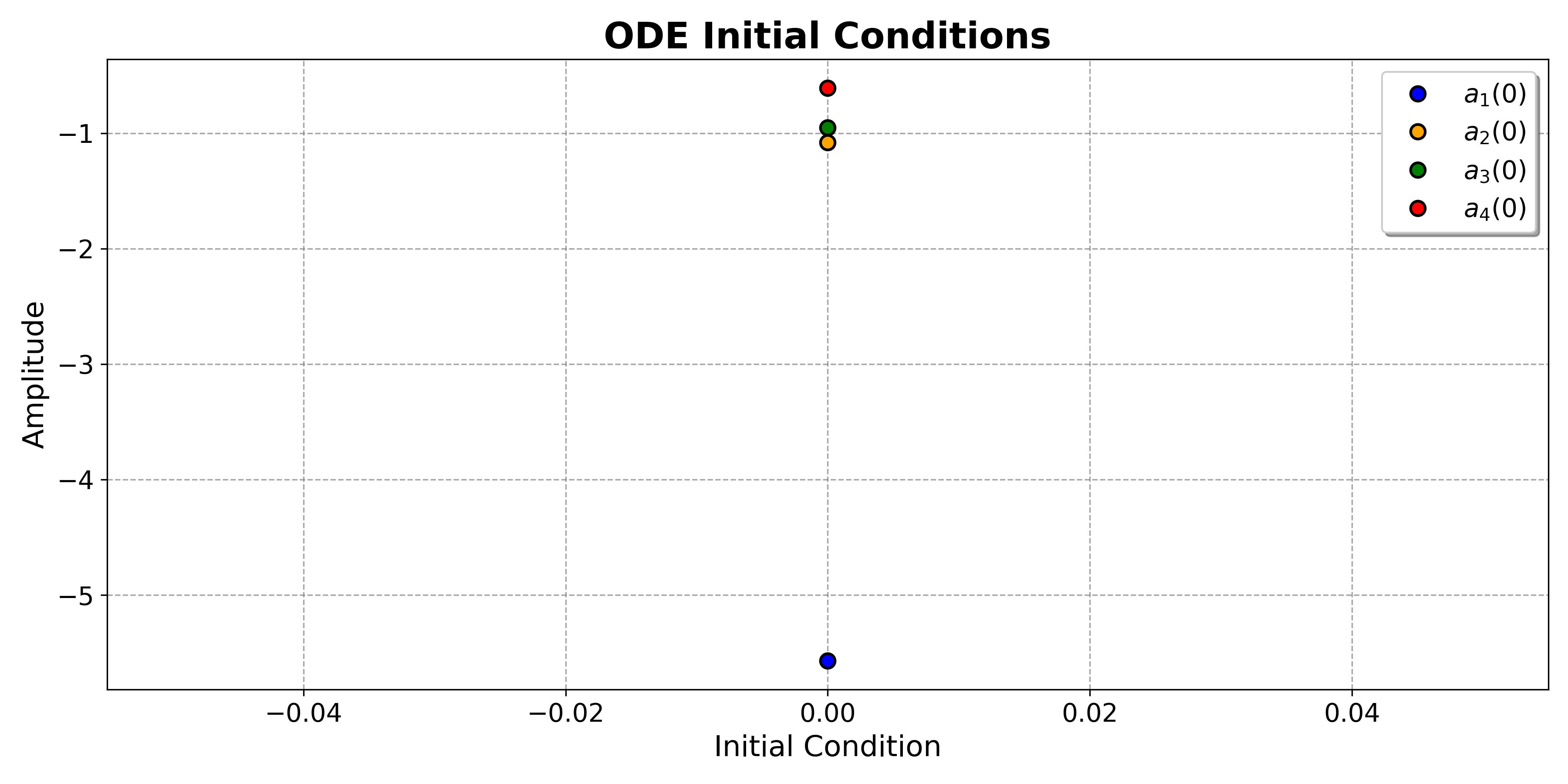}
         \caption{}
     \end{subfigure}
     \hfill
     \begin{subfigure}[b]{0.45\textwidth}
         \centering
         \includegraphics[scale=0.225]{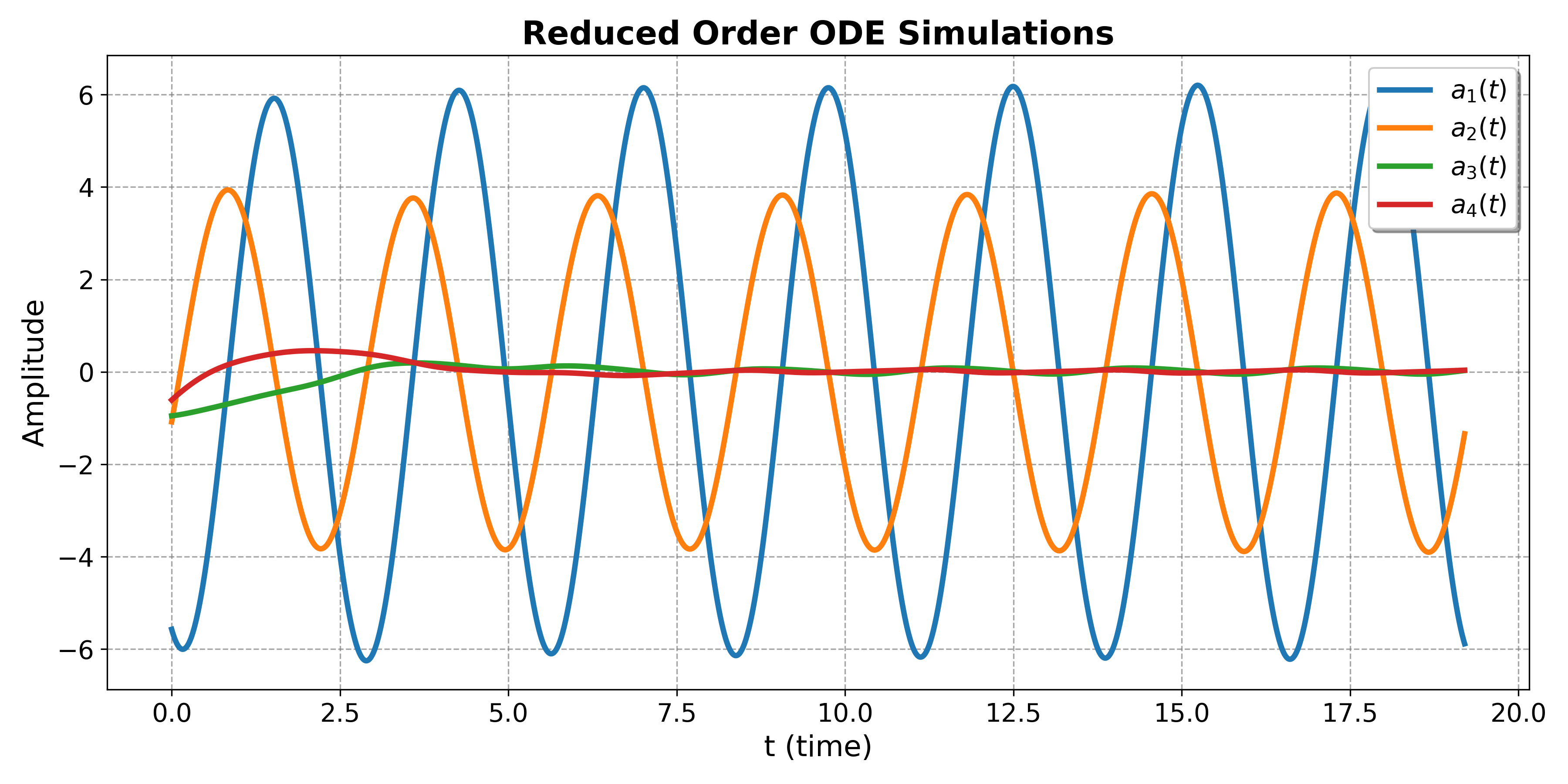}
         \caption{}
     \end{subfigure}
      \begin{subfigure}[c]{0.6\textwidth}
         \centering
         \includegraphics[scale=0.3]{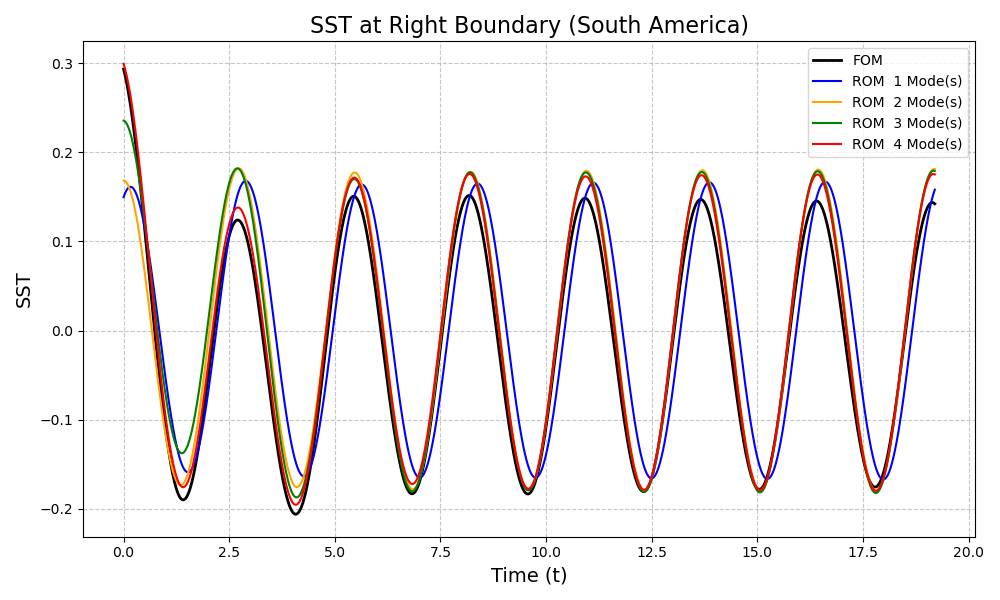}
         \caption{}
     \end{subfigure}
     \caption{(a) Projected initial conditions  (b) reduced order equations $a_{\alpha}(t)$  from POD-Galerkin projection and (c) reconstruction of SST time series at the right boundary with different number of modes}
      \label{fig:POD_Galerkin_1}
\end{figure}     
\textbf{Remark 4.2 :} Using this approach, we emphasize that  each component of ENSO model \eqref{eq:Coupled_Model} has distinct spatial modes written as a vector form in $\{ \hat{\Psi}_{\alpha}(x) \}_{\alpha=1}^n$ and $\{ \tilde{\Psi}_{\alpha}(x) \}_{\alpha=1}^n$, reflecting their unique variability patterns. However, the temporal coefficients $\{a_{\alpha} (t)\}_{\alpha=1}^n$ remain the same across all components, ensuring a unified low-dimensional representation of the system’s dynamics as given in Equation \eqref{eq:Reduced_ODEs}. That approach makes it uniquely different from other applications of the POD-Galerkin method. \\

Figure \eqref{fig:POD_Galerkin_1} depicts the projected initial conditions and the simulation of the reduced order equations using these conditions. The solutions for the ODEs, $a_{\alpha}(t)$, and POD time coefficients, $q_{\alpha} (t),$, where $\alpha =1, \dots ,4$, are very similar demonstrating the validity of the Galerkin projection. The reconstruction L1 errors of the SST time series at the right boundary (South America) using different number of POD modes and time coefficients $\alpha(t)$ are  0.04713 (n=1),  0.02375 (n=2),  0.02192 (n=3), and 0.01518 (n=4), respectively. These results demonstrate an improvement in reconstruction accuracy as the number of modes increases. This reflects the efficiency of POD-Galerkin in capturing dominant dynamics with fewer modes, as shown in Figure \eqref{fig:POD_Galerkin_1}. The decreasing error highlights the method's capability to approximate complex systems with reduced computational effort.

\section{Results}\label{sc:results}
 In this section, we demonstrate the accuracy of the proposed methods using reconstruction errors relative to the full-order model. The reconstruction of model components given in Equation \eqref{eq:solnbar} can be performed in two ways: (i) using POD, where the approximation with four modes $\tilde{\Psi}_{\alpha}(x)$ and time coefficients $q_{\alpha}(t)$ is given by
\begin{equation}
    \overline{u}_{\text{POD}}(t,x) \approx \sum_{\alpha=1}^4 q_{\alpha} (t) \tilde{\Psi}_{\alpha}(x) \label{eq:POD_approx}
\end{equation}
and (ii) using POD-Galerkin with POD modes $\tilde{\Psi}_{\alpha}(x)$ and ODEs $a_{\alpha}(t)$, where the reconstruction takes the form 
\begin{equation}
    \overline{u}_{\text{POD-G}}(t,x) \approx \sum_{\alpha=1}^4 a_{\alpha} (t) \tilde{\Psi}_{\alpha}(x) \label{eq:POD-G_approx}
\end{equation}

\begin{figure}[H]
    \centering
    \includegraphics[scale=0.45]{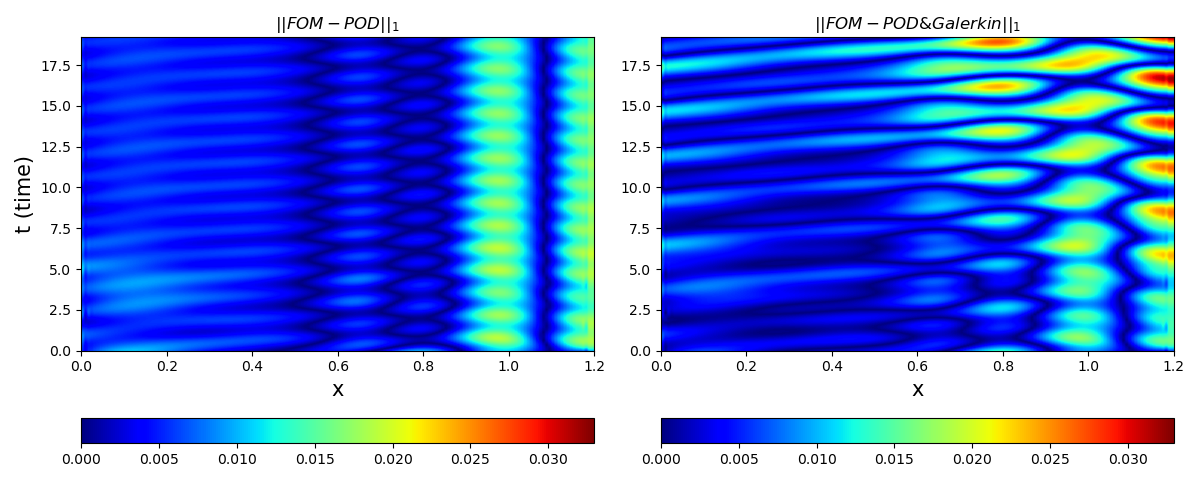}
    \caption{Reconstruction errors ($L1$) of full order SST component $T(t,x)$ using POD and POD-Galerkin methods with four modes}
    \label{fig:POD_G_Temp_Reconstruction_2D_error}
\end{figure}

The first method involves POD modes $\tilde{\Psi}(x)$ and time coefficients $q_{\alpha}(t)$ \textit{without} any reduced-order equations as shown in Figure \eqref{fig:POD1}. On the other hand, the second method incorporates the simulation of ODEs, $a_{\alpha}(t)$, as depicted in Figure \eqref{fig:POD_Galerkin_1} where the coefficients for the equations are obtained with the Galerkin projection using POD modes. In both cases, while POD modes $\tilde{\Psi}_{\alpha}(x)$ are same, the time coefficients $a_{\alpha}(t)$ and $q_{\alpha}(t)$ are slightly different. However, both approximations are highly accurate using four time coefficients and four POD modes. 
\begin{figure}[H]
    \centering
    \includegraphics[scale=0.45]{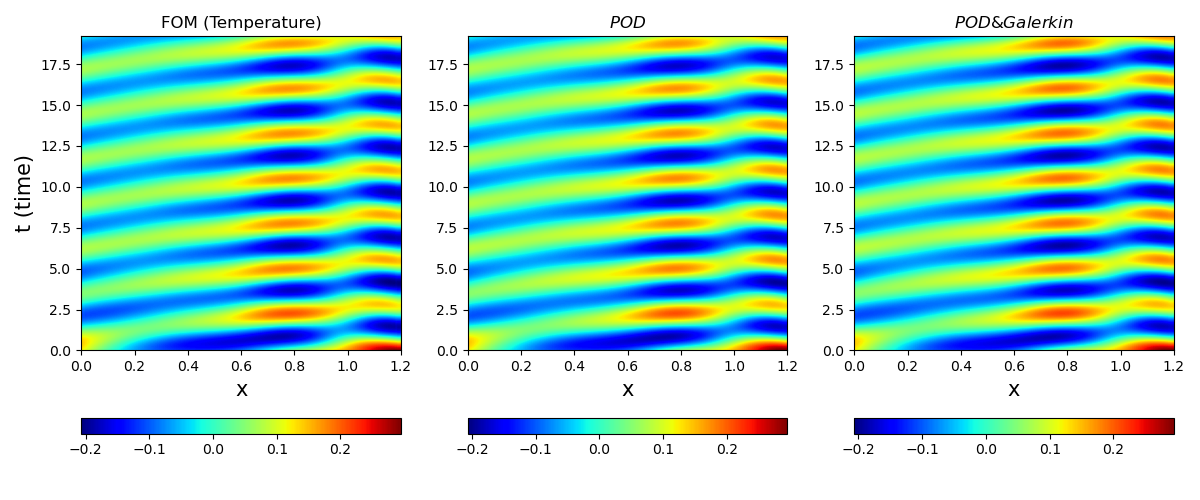}
    \caption{Full order model (FOM), POD and POD-Galerkin (POD-G) reconstructions of SST component $T(t,x)$ using four modes}
    \label{fig:POD_G_Temp_Reconstruction_2D}
\end{figure}   
The accuracy of these methods using four time coefficients and four POD modes are illustrated in Figure \eqref{fig:POD_G_Temp_Reconstruction_2D} and \eqref{fig:POD_G_Temp_Reconstruction_2D_error} for the SST model $T(t,x)$. Figure \eqref{fig:POD_G_Temp_Reconstruction_2D} (left)  depicts the simulation of full order model of SST $T(t,x)$ using finite difference methods. The reconstruction using $\overline{u}_{\text{POD}}(t,x)$  in Equation \eqref{eq:POD_approx} and POD-G $\overline{u}_{\text{POD-G}}(t,x)$  in Equation \eqref{eq:POD-G_approx} are shown in Figure \eqref{fig:POD_G_Temp_Reconstruction_2D} (middle and right), respectively. Both methods achieve over 95\% accuracy, although POD reduction has a lower L1 error using $|u(t,x) - \overline{u}_{\text{POD}}(t,x)|$ as shown in Figure \eqref{fig:POD_G_Temp_Reconstruction_2D_error} (left). However, while POD combined with Galerkin has a slightly larger error $|u(t,x) - \overline{u}_{\text{POD-G}}(t,x)|$, it provides analytical capabilities via reduced-order ODEs. The main drawback of the POD approximation without Galerkin projection is that it does not provide model equations for analyzing system dynamics for further applications. However, although POD-Galerkin reduction is slightly less accurate (Figure \eqref{fig:POD_G_Temp_Reconstruction_2D_error}, right), it provides a set of ODEs for reduced-order model analysis, such as bifurcation analysis and SST estimation.

\section{Conclusion and Future Work}\label{sc:conc}
 This study has demonstrated the efficacy of the POD-Galerkin projection in modeling the El Niño Southern Oscillation (ENSO) phenomena. By capturing the full order model with only four modes and four reduced order equations, we achieved a substantial reduction in computational complexity without significant loss of accuracy. Due to the special properties of the model, we introduced a novel approach using different POD bases, but same time coefficients for all model components. \\

Our results illustrated the capability of the reduced-order model to replicate key oceanographic features associated with ENSO, including oceanic Kelvin waves, oceanic Rossby waves, and variations in sea surface temperature (SST). The reduced-order model was able to maintain the essential dynamics and spatiotemporal patterns of these features, which are critical for understanding and predicting ENSO events. Overall, the application of the POD-Galerkin method to ENSO modeling presents a promising approach for efficient and accurate simulations. This reduction in model complexity could facilitate more extensive and rapid assessments of climate scenarios, potentially enhancing predictive capabilities. \\

Future work will concentrate on the reduced-order modeling of stochastic coupled ENSO models and assessing the efficiency of the proposed approach. This study lays the groundwork for developing reduced-order filtering techniques,that is a data assimilation technique that integrates observational data into reduced-order models to estimate and update the state of complex dynamical systems efficiently. By combining dimensionality reduction methods with state estimation techniques, like particle filters, reduced-order filtering retains essential dynamics while significantly reducing computational costs. This approach is vital for enhancing the accuracy and efficiency of climate models, enabling better predictions of ENSO.

\begin{table}[H]
\centering
\caption{Descriptions of parameters in the operator $\A(\kappa)$ and nonlinearity $\N(u,u,\mu)$.}
\label{tab:parameters}
\begin{tabular}{|c|l|c|}
\hline
\textbf{Parameter} & \textbf{Description} & \textbf{Nondimensional Values} \\ \hline
$\kappa$ & Wind stress coefficient & 5.6 \\ \hline
$\gamma$ & Atmospheric damping coefficient & 0.1 \\ \hline
$\chi_{A}$ & Meridional projection coefficient from ocean to atmosphere & 0.31 \\ \hline
$\chi_{O}$ & Meridional projection coefficient from atmosphere to ocean & 1.38 \\ \hline
$\alpha_{q}$ & Latent heating factor & 0.2 \\ \hline
$\overline{Q}$ & Mean state of heat flux & 0.9 \\ \hline
$c$ & Wave propagation speed & 0.5 \\ \hline
$\delta$ & Oceanic damping coefficient & 0.5 \\ \hline
$\eta(x)$ & Zonal thermocline feedback profile & $1.5 + 0.5 \times \tanh(7.5x - L_O/2)$ \\ \hline
$\xi$ & Coupling strength between ocean and atmosphere & 8.5 \\ \hline
$\mu$ & Nonlinear zonal advection coefficient & 0.04 \\ \hline
$r_W$ & Western boundary reflection coefficient in ocean & 0.5 \\ \hline
$r_E$ &  Eastern boundary reflection coefficient in ocean & 0.5 \\ \hline
$L_A$ &  Equatorial belt length & 2.6 \\ \hline
$L_O$ &  Equatorial Pacific length & 1.2 \\ \hline
\end{tabular}
\end{table}

\section*{Data Accessibility}
The datasets and code used to produce the results in this paper can be accessed at \url{https://github.com/Y1Aydogdu/ENSO_POD_Galerkin.git}

\section*{Acknowledgments} The authors acknowledge partial support for this work from Natural Sciences and Engineering Research Council (NSERC) Discovery grant 50503-10802, TECSIS/ Fields-CQAM Laboratory for Inference and Prediction and NSERC-CRD grant 543433-19. The authors would like to thank Professor Peter Baxendale, Department of Mathematics, University of Southern California, for stimulating discussions, contributions and suggestions on many topics of this paper.


\section{Appendix A}
We thank Peter Baxendale for permission to copy the material in \cite{Baxendale2024}.
\subsection{Proof of Theorem \ref{thm opt}}

For ease of notation write $w(t_i) = w_i$ for $1 \le i \le M$.

 \subsection{Operators $\Gamma$ and $\Gamma^\ast$}  Let $\langle \cdot,\cdot \rangle_M$ denote the usual Euclidean inner product on $\R^M$.  
Given the snapshots $\{w_i \in H: 1 \le i \le M\}$, define the operator $\Gamma: H \to \R^M$ by 
    \begin{equation} \label{G}
    (\Gamma \phi)_i = \langle \phi, w_i \rangle_H, \qquad 1 \le i \le M.
    \end{equation} 
If $\phi \in H$ and $g \in \R^M$ then 
$$
     \langle \Gamma \phi, g \rangle_M   =  \sum_{i=1}^M \langle \phi, w_i\rangle_H g_i  =  \langle \phi, \Gamma^{\ast} g \rangle_H
    $$
where $\Gamma^\ast: \R^M \to H$ is given by 
    \begin{equation}\nonumber
    \Gamma^\ast g =   \sum_{i=1}^M g_i w_i.
    \end{equation}

\begin{lemma} \label{lem trace} Let $\{\phi^\alpha: \alpha \ge 1\}$ be any complete orthonormal basis for $H$.  Then
    \begin{equation} \label{trace}
   \sum_{\alpha \ge 1} \langle \phi^\alpha,\Gamma^\ast \Gamma \phi^\alpha\rangle_H =  \sum_{i=1}^M  \|w_i\|_H^2 < \infty.
    \end{equation}
\end{lemma}
    
\n{\bf Proof.}  We have
    $$
    \sum_{\alpha \ge 1} \langle \phi^\alpha,\Gamma^\ast \Gamma \phi^\alpha\rangle_H
     = \sum_{\alpha \ge 1} \left(   \sum_{i=1}^M\langle \phi^\alpha,w_i \rangle_H^2 \right)
          =  \sum_{i=1}^M \left(\sum_{\alpha \ge 1} \langle \phi^\alpha,w_i \rangle_H^2\right) 
         =   \sum_{i=1}^M \|w_i\|_H^2 .
  $$
\qed

\subsection{Changing the problem}  For $n \ge 1$ define
   \begin{equation} \label{muhat}
    \widehat{\mu}_n = \sup\left\{ \sum_{\alpha = 1}^n \langle \Gamma^\ast \Gamma \psi^\alpha,\psi^\alpha \rangle_H: \mbox{ the set }\{\psi^1,\ldots,\psi^n \} \mbox{ is orthonormal in }H\right\}.
    \end{equation}
   The following result is closely based on the essential idea in Theorem 1 in \cite{Djouadi2012}.  

\begin{theorem}  \label{thm muhat mu} Suppose the supremum in the maximization problem \eqref{muhat} is attained by an orthonormal set $\{\widehat{\psi}^1,\ldots,\widehat{\psi}^n\}$.   Then the minimization problem \eqref{mu} has $\mu_n = \sum_{i=1}^M\|w_i\|_H^2 - \widehat{\mu}_n$ and the infimum is attained with $F_i = \sum_{\alpha=1}^n (\Gamma \widehat{\psi}^\alpha)_i \widehat{\psi}^\alpha$.
\end{theorem}

\n{\bf Proof.}  Write $F_i = \sum_{\alpha = 1}^n a^\alpha_i \psi^\alpha$.  The Gram-Schmidt orthogonalization process yields a complete orthonormal system $\{\phi^\alpha:  \alpha \ge 1\}$ such that $\psi^\alpha \in \mbox{span}\{\phi^1,\ldots,\phi^n\}$ for all $\alpha = 1, \ldots, n$.  Thus, after some elementary rearranging, we may assume $F$ is of the form $\sum_{\alpha = 1}^n b^\alpha_i \phi^\alpha$ where the set $\{\phi^1,\ldots\phi^n\}$ gives the first $n$ terms of a complete orthonormal system $\{\phi^\alpha: \alpha \ge 1\}$.  
Then
  \begin{align*}
   \sum_{i=1}^M \|w_i - F_i\|_H^2 & = \sum_{i=1}^M \Bigl\|w_i - \sum_{\alpha = 1}^n b^\alpha_i\phi^\alpha \Bigr\|_H^2\\
          & =  \sum_{i=1}^M  \sum_{\beta \ge 1} \Bigl\langle \phi^\beta, \Bigl(w_i - \sum_{\alpha= 1}^n b^\alpha_i \phi^\alpha\Bigr) \Bigr\rangle_H^2\\
          & = \sum_{\beta \ge 1} \sum_{i=1}^M  \Bigl( (\Gamma \phi^\beta)_i -  \sum_{\alpha = 1}^n b^\alpha_i\langle \phi^\beta,\phi^\alpha\rangle_H\Bigr)^2 \\
   & = \sum_{\beta=1}^n \sum_{i=1}^M \Bigl( (\Gamma \phi^\beta)_i -  b^\beta_i \Bigr)^2 + \sum_{\beta > n} \sum_{i=1}^M \Bigl( (\Gamma \phi^\beta)_i \Bigr)^2 \\
   & = \sum_{\beta=1}^n \sum_{i=1}^M  \Bigl( (\Gamma \phi^\beta)_i -  b^\beta_i \Bigr)^2dt  + \sum_{i=1}^M \|w_i\|_H^2   -   \sum_{\beta =1}^n \langle \Gamma^\ast \Gamma \phi^\beta, \phi^\beta \rangle_H.
   \end{align*}
where the last equality uses 
\eqref{trace}. 
The first term can be minimized by choosing $b^\beta_i = (\Gamma \phi^\beta)_i$ for $1 \le \beta \le n$ and the third term can be minimized by using the solution of Problem 2.  Thus $\mu_n = \sum_{i=1}^M \|w_i\|_H^2 - \widehat{\mu}_n$ and the optimizing function is given by $\phi^\alpha(x) = \widehat{\psi}^\alpha(x)$ and then $b^\alpha = (\Gamma \widehat{\psi}^\alpha)$.  \qed

\subsection{Eigenvalues and eigenfunctions of $\Gamma^\ast \Gamma$}  

The operator $\Gamma^\ast \Gamma$ maps $H$ into itself.  It is symmetric and non-negative and of rank at most $M$.  Thus there is a complete orthonormal set of eigenfunctions $\{\nu^\alpha: \alpha \ge 1\}$ in $H$ and corresponding eigenvalues $\widehat{\lambda}_1 \ge \widehat{\lambda}_2 \ge \widehat{\lambda}_3 \ge \cdots \ge 0$ (counted with multiplicities) so that 
     $$
     (\Gamma^\ast \Gamma)\nu^\alpha = \widehat{\lambda}^\alpha \nu^\alpha,  \quad \alpha \ge 1.
     $$
 (Since $H$ is a space of functions, we refer here to eigenvectors of $\Gamma^\ast \Gamma$ as eigenfunctions.)

\begin{theorem} \label{thm muhat} For the maximization problem \eqref{muhat}, we have $\widehat{\mu}_n = \sum_{\alpha = 1}^n \widehat{\lambda}_\alpha$, attained by the orthonormal set $\{\nu^{1}, \ldots,\nu^n\}$.
\end{theorem} 

\n{\bf Proof.}  Let $\{\psi^1, \ldots,\psi^n\}$ be an orthonormal set in $H$, and define the $n \times n$ matrix $A$ by $A_{\alpha \beta} = \langle \Gamma^\ast  \Gamma \psi^\alpha,\psi^\beta \rangle$.  Let $\kappa_1 \ge \kappa_2 \ge \cdots \ge \kappa_r > 0$ denote the strictly positive eigenvalues of $A$.  The Rayleigh-Ritz procedure applied to $-\Gamma^\ast \Gamma$ and $-A$, see for example Theorem 10.12 in \cite{Nakao2019}, implies $\widehat{\lambda}_\alpha \ge \kappa_\alpha$ for $1 \le \alpha \le r$.  Then
   $$
   \sum_{\alpha = 1}^n  \langle \Gamma^\ast  \Gamma \psi^\alpha,\psi^\alpha \rangle_H
   = \sum_{\alpha = 1}^n A_{\alpha \alpha} = \mbox{trace}(A) = \sum_{\alpha =1}^n \kappa_\alpha  \le \sum_{\alpha = 1}^n \widehat{\lambda}_\alpha.
   $$
Therefore $\widehat{\mu} \le \sum_{\alpha = 1}^n \widehat{\lambda}_\alpha$, and equality is attained by choosing $\psi^\alpha = \nu^\alpha$.  \qed 

\s

Applying the result of Theorem \ref{thm muhat} in Theorem \ref{thm muhat mu} and noting
 $$ \sum_{\alpha \ge 1} \widehat{\lambda}^\alpha =  \sum_{\alpha \ge 1} \langle \nu^\alpha, (\Gamma^\ast \Gamma)\nu^\alpha \rangle_H = \sum_{i=1}^M \|w_i\|_H^2,
 $$
we get:

\begin{corollary} \label{cor Gamma} Suppose $\Gamma^\ast \Gamma$ has eigenvalues $\widehat{\lambda}_1 \ge \widehat{\lambda}_2 \ge \cdots \ge 0$ (counted with multiplicities) with orthonormal eigenfunctions $\nu^1, \nu^2, \ldots$.  Then for $n \ge 1$, the minimization problem \eqref{mu} has solution $\mu_n = \sum_{\alpha > n} \widehat{\lambda}_\alpha$ attained with $F_i = \sum_{\alpha=1}^n (\Gamma \nu^\alpha)_i \nu^\alpha$.
\end{corollary}    

\begin{remark}  Suppose that instead of snapshots we have the full continuous time observations $\{w(t) \in H: 0 \le t \le T\}$.  In this setting it is natural to consider 
   \begin{equation} \label{mu cont}
    \mu_n = \inf_{F \in {\cal S}_n} \left( \int_0^T \|w(t)- F(t)\|_H^2\right)
   \end{equation}
where now ${\cal S}_n$ is the set of all functions $F:[0,T] \to H$ of the form 
   $$
   F(t) = \sum_{\alpha= 1}^n a^\alpha(t)\psi_\alpha
   $$
where all the functions $a^\alpha$ are in $L^2([0,T])$ and all the $\psi_\alpha$ are in $H$.  Define $\Gamma: H \to L^2([0,T])$ by $(\Gamma \phi)(t) = \langle \phi, w(t)\rangle_H.$  The above calculations remain essentially unchanged when $\R^M$ is replaced with $L^2([0,T])$.  Note especially that the argument in Lemma \ref{lem trace} extends easily to the continuous case so that $\Gamma^\ast \Gamma$ is trace class and hence compact, even though it is no longer guaranteed to be of finite rank.  It follows that the natural extension of Corollary \ref{cor Gamma} to the continuous time minimization problem \eqref{mu cont} is valid.  

\end{remark}

\subsection{Eigenvalues and eigenvectors of $C$}  

Corollary \ref{cor Gamma} gives the solution of the minimization problem \eqref{mu} in terms of the operator $\Gamma^\ast \Gamma$ on the Hilbert space $H$.  Here we convert it to the computationally useful form stated in Theorem \ref{thm opt} in terms of the $M \times M$ matrix $C$. Notice that for $\gamma \in \R^M$ we have $(C\gamma)_i = \sum_{j=1}^M \langle w_i,w_j \rangle \gamma_j = \langle w_i, \Gamma^\ast \gamma \rangle  = (\Gamma \Gamma^\ast \gamma)_i$, so that $C = \Gamma \Gamma^\ast$. . 

\begin{proposition}  \label{prop GC2}  (i)  If $\{\phi^1,\ldots,\phi^n\}$ is any orthogonal set of non-zero eigenfunctions of $\Gamma^\ast \Gamma$ with non-zero eigenvalues $\widetilde{\lambda}_1, \ldots \widetilde{\lambda}_n$, then $\{\Gamma\phi^1,\ldots,\Gamma\phi^n\}$ is an orthogonal set of non-zero eigenvectors of $C$ with the same non-zero eigenvalues $\widetilde{\lambda}_1, \ldots \widetilde{\lambda}_n$.

(ii) Conversely, if $\{\gamma^1,\ldots,\gamma^n\}$ is any orthogonal set of non-zero eigenvectors of $C$ with non-zero eigenvalues $\widetilde{\lambda}_1, \ldots \widetilde{\lambda}_n$, then $\{\Gamma^\ast\gamma^1,\ldots,\Gamma^\ast\gamma^n\}$ is an orthogonal set of non-zero eigenfunctions of $\Gamma^\ast \Gamma$ with the same non-zero eigenvalues $\widetilde{\lambda}_1, \ldots \widetilde{\lambda}_n$.

\end{proposition}

\n{\bf Proof.} (i) We have
   $$
   C(\Gamma \phi^\alpha) = \Gamma \Gamma^\ast \Gamma \phi^\alpha = \Gamma \bigl(\Gamma^\ast \Gamma \phi^\alpha\bigr) = \Gamma\bigl( \widetilde{\lambda}_\alpha \phi^\alpha\bigr) = \widetilde{\lambda}_\alpha \bigl( \Gamma \phi^\alpha \bigr)
   $$
and 
  $$
  \langle \Gamma \phi^\alpha, \Gamma \phi^\beta \rangle_M =  \langle \phi^\alpha, \Gamma^\ast \Gamma \phi^\beta \rangle_H = \widetilde{\lambda}_\beta \langle \phi^\alpha,\phi^\beta \rangle _H.
    $$
Together, we see $\Gamma \phi^\alpha \neq 0$ and the $\Gamma \phi^1,\ldots, \Gamma \phi^n$ are orthogonal and $\Gamma \phi^\alpha$ is an eigenvector of $C$ with the same eigenvalue $\widetilde{\lambda}_\alpha$ as the eigenfunction $\phi^\alpha$.

(ii)  Similarly we have 
  $$
   \Gamma^\ast \Gamma(\Gamma^\ast \gamma^\alpha) = \Gamma^\ast\bigl(C \gamma^\alpha\bigr) = \Gamma^\ast\bigl( \widetilde{\lambda}_\alpha \gamma^\alpha\bigr) = \widetilde{\lambda}_\alpha \bigl( \Gamma^\ast \gamma^\alpha \bigr)
   $$
and 
  \begin{equation} \label{normalize}
  \langle \Gamma^\ast \gamma^\alpha, \Gamma^\ast \gamma^\beta \rangle_H =  \langle \gamma^\alpha, \Gamma \Gamma^\ast \gamma^\beta \rangle_M = \langle \gamma^\alpha, C \gamma^\beta \rangle_M  = \widetilde{\lambda}_\beta \langle \gamma^\alpha,\gamma^\beta \rangle _M,
    \end{equation}
and the result follows as in (i).  \qed

\s

\n{\bf Proof of Theorem \ref{thm opt}}. Let $\lambda_\alpha$ and $\gamma^\alpha$ be the eigenvalues and orthonormal eigenvectors of $C$, as in the statement of Theorem \ref{thm opt}.  Then in Corollary \ref{cor Gamma} we have $\widehat{\lambda}_\alpha = \lambda_\alpha$ and orthogonal eigenfunction $\Gamma^\ast \gamma^\alpha$.  Using \eqref{normalize} we have orthonormal eigenvectors
   $$
   \psi_\alpha = \frac{1}{\sqrt{\lambda_\alpha}}(\Gamma^\ast \gamma^\alpha) = \frac{1}{\sqrt{\lambda_\alpha}} \sum_{i=1}^M \gamma_i^\alpha w_i
   $$
and then 
    $$
    a^\alpha_i = (\Gamma \psi_\alpha)_i =  \frac{1}{\sqrt{\lambda_\alpha}}(\Gamma\Gamma^\ast \gamma^\alpha) = \sqrt{\lambda_\alpha} \gamma^\alpha_i.
   $$
   \qed
\section{Appendix B}
It is a feature of the coupled ENSO model~\eqref{eq:Coupled_Model} that there
are no time derivatives of the first two components, $K^A(t,x)$ and $R^A(t,x)$. Hence, the truncated atmospheric model can be explicitly written as,
\begin{equation}\label{trun-atmosphere}
\begin{aligned}
    \partial_{x} K^A(t,x)  &= \chi_{A} (2 - 2 \overline{Q})^{-1}\alpha_{q} T(t,x)   - \gamma K^A(t,x),    \quad 0\leq x \leq L_O\\
     K^A(t,0)&=e^{-\gamma( L_A- L_O)} K^A(t,L_O)\\
     -\partial_{x} R^A(t,x) &= 3\chi_{A}  (3 -  3 \overline{Q})^{-1} \alpha_{q} T(t,x)  - 3\gamma R^A(t,x),   \quad 0\leq x \leq L_O\\
     R^A(t,0)&=e^{3 \gamma( L_A- L_O)} R^A(t,L_O).
\end{aligned}
\end{equation}
As discussed in Zebiak~\cite{CZ1987} the parameter  $\gamma$ is intended to represent the effect of a wide
range of physical processes and as such is not easy to determine a priori. Gill~\cite{Gill1980} uses a value
of $0.1$, while Zebiak~\cite{CZ1987} uses a value of $0.3$ arguing that that value gave, in his case, good agreement
between model results and observations. A wide range of values of  $\gamma$ may be used without
affecting qualitative conclusions. 
In the first order boundary value problems~\eqref{trun-atmosphere}, $t$ is a parameter and $T(t,x)$ is given.
The general solutions for $K^A(t,x)$ and $R^A(t,x)$ are given explicitly in terms of $T(t,x)$ as
\begin{equation}
\label{eq:atmosM-truncate-Ksol}
\begin{aligned}
K^A(t,x)  &=\frac{e^{-\gamma x}\,\chi_{A} (2 - 2 \overline{Q})^{-1}\,\alpha_{q}\, e^{-\gamma L_A}}{1-e^{-\gamma L_A}}\int_{0}^{L_O} e^{\gamma\xi}T(t,\xi) d\xi \\
      &\hspace*{2.0cm}+\chi_{A} (2 - 2 \overline{Q})^{-1}\alpha_{q} e^{-\gamma x}\int_{0}^{x} e^{\gamma\xi}T(t,\xi) d\xi, \quad 0\leq x \leq L_O,
\end{aligned}
\end{equation}
and
\begin{equation}
\label{eq:atmosM-truncate-Rsol}
\begin{aligned}
R^A(t,x)  &=\frac{-e^{3\gamma(x-L_{O})}3\chi_{A} (3 - 3 \overline{Q})^{-1}\,\alpha_{q}\, e^{-3\gamma L_A}}{1-e^{-3\gamma L_A}}\int_{0}^{L_O} e^{-3\gamma (\xi-L_O)}T(t,\xi) d\xi  \\
      &\hspace*{1.5cm}-3\chi_{A} (3 - 3 \overline{Q})^{-1}\alpha_{q} e^{3\gamma(x-L_{O})}\int_{x}^{L_{O}} e^{-3\gamma}(\xi-L_{O})T(t,\xi) d\xi, \quad 0\leq x \leq L_O.
\end{aligned}
\end{equation}

\end{document}